\documentclass[letterpaper,twocolumn,10pt]{article}
\usepackage{usenix}

\usepackage{amsmath}
\usepackage{xcolor}
\usepackage{xspace}
\usepackage{graphicx}
\usepackage{booktabs}
\usepackage{tabularx}
\usepackage{makecell}
\usepackage{rotating}
\usepackage{tablefootnote}
\usepackage{listings}
\usepackage{xurl}
\usepackage{hyperref}
\usepackage{appendix}

\usepackage{booktabs}
\usepackage{graphicx}
\usepackage[nolist]{acronym}
\usepackage{listings}
\usepackage{algorithm}
\usepackage{algorithmicx}
\usepackage{algpseudocode}
\usepackage[binary-units,abbreviations]{siunitx}
\usepackage{amsmath}
\usepackage{pgfplots}
\usepackage{tikzscale}
\usepackage{wasysym}
\usepackage{enumitem}
\usepackage{amsmath, amssymb, amsthm}
\usepackage[USenglish]{babel}

\AtBeginDocument{%
  \providecommand\BibTeX{{%
    \normalfont B\kern-0.5em{\scshape i\kern-0.25em b}\kern-0.8em\TeX}}}

\usepackage{ifthen}

\usepackage{booktabs}
\usepackage{pifont}

\lstdefinestyle{mystyle}{
    numberstyle=\tiny,
    basicstyle=\ttfamily\footnotesize,
    breakatwhitespace=false,         
    breaklines=true,                 
    captionpos=b,                    
    keepspaces=true,                 
    numbers=left,                    
    numbersep=5pt,                  
    showspaces=false,                
    showstringspaces=false,
    showtabs=false,                  
    tabsize=2
}

\begin{document}

\date{}
\title{\Large \bf Insecure Despite Proven Updated: Extracting the Root VCEK Seed on EPYC Milan via a Software-Only Attack}

\author{
{
\rm Muyan Shen$^{\dagger\ddagger}$, Yu Qin$^{\ddagger}$}\\
\textit{$^{\dagger}$University of Chinese Academy of Sciences}\\
\textit{$^{\ddagger}$Institute of Software, Chinese Academy of Sciences}
} %

\maketitle

\begin{abstract}
In the official whitepaper of \ac{sev-snp}, AMD explicitly emphasizes the capability to prevent Trusted Computing Base (TCB) rollback attacks. Cryptographically, this is realized by signing attestation reports with the \ac{vcek}, which is derived by incorporating the TCB version into the hardware root seed.

In this architecture, safeguarding the hardware root seed is the ultimate line of defense. However, our research reveals that this protection is insufficient on EPYC Milan by presenting a software-only exploit. Specifically, we firstly introduce MilanLaunchy attack, an exploit that achieves code execution on the AMD secure processor. Building on this foundation, we develop the BadFuse attack, which extracts the hardware root seed by exploiting a lack of write restrictions in the fuse controller. 
This end-to-end attack chain enables an adversary to forge valid attestation reports for any firmware version, thereby effectively undermining the security model of SEV-SNP.

\end{abstract}

\begin{acronym}
    \acro{dice}[DICE]{Device Identifier Composition Engine}
    \acro{umc}[UMC]{Unified Memory Controller}
    \acro{abl}[ABL]{AGESA Boot Loader}
    \acro{kds}[KDS]{Key Distribution Service}
    \acro{vm}[VM]{virtual machine}
    \acro{abi}[ABI]{Application Binary Interface}
    \acro{tdx}[TDX]{Trust Domain Extensions}
    \acro{tee}[TEE]{Trusted Execution Environment}
    \acro{cc}[CC]{Confidential Computing}
    \acro{amd-sp}[ASP]{AMD Secure Processor}
    \acro{smu}[SMU]{System Management Unit}
    \acro{sev}[SEV]{Secure Encrypted Virtualization}
    \acro{vm}[VM]{virtual machine}
    \acro{sev-es}[SEV-ES]{SEV Encrypted State}
    \acro{sev-snp}[SEV-SNP]{Secure Encrypted Virtualization with Secure Nested Paging}
    \acro{csp}[CSP]{Cloud Service Provider}
    \acro{cek}[CEK]{Chip Endorsement Key}
    \acro{vcek}[VCEK]{Versioned Chip Endorsement Key}
    \acro{vlek}[VLEK]{Versioned Loaded Endorsement Key}
    \acro{tcb}[TCB]{Trusted Computing Base}
    \acro{rom}[ROM]{read-only memory}
    \acro{svn}[SVN]{security version number}
    \acro{ark}[ARK]{AMD Root Key}
    \acro{bmc}[BMC]{Baseboard Management Controller}
    \acro{spd}[SPD]{Serial Presence Detect}
\end{acronym}

\acresetall
\section{Introduction\label{sec:intro}}

The paradigm of confidential computing (CC) relies on trusted execution environments (TEEs) to protect data in use from completely untrusted hypervisors. Leading architectures, such as AMD \ac{sev-snp}\cite{sev2020strengthening}, Intel \ac{tdx}\cite{Intel_TDX}, and ARM Confidential Compute Architecture (CCA)\cite{ARM_CCA}, all leverage robust hardware-backed isolation to shield guest virtual machines (VMs). Within these diverse ecosystems, remote attestation serves as the foundational pillar of trust. This cryptographic mechanism allows cloud tenants to verify attestation reports, enabling them to irrefutably confirm the initial state of their VMs and the underlying platform's firmware security.

Within AMD SEV-SNP, the architecture delegates attestation reporting to the AMD Secure Processor (ASP)\cite{ASP_coreboot,psptool-github,specter-2024-part1}, an embedded ARM core executing with the highest system privileges. The ASP firmware constitutes a fundamental component of the SEV-SNP Trusted Computing Base (TCB) and is a primary focus of remote attestation.
Crucially, it is imperative that the firmware version reported in the attestation cannot be misrepresented or spoofed, even if an adversary achieves arbitrary code execution on an outdated/vulnerable ASP firmware version.
Consequently, AMD explicitly emphasizes the architecture's ability to defend against TCB rollback attacks in the official SEV-SNP whitepaper\cite{sev2020strengthening}, highlighting the prevention of reverting the ASP firmware to older versions, which is a critical protection notably absent in its predecessors, \ac{sev} and \ac{sev-es}\cite{AMDSevAPI}.

To cryptographically enforce this anti-rollback policy, AMD fundamentally redesigned the hardware root of trust. In earlier SEV iterations, attestation reports were signed using a static \ac{cek}. Since the CEK did not evolve with firmware updates, versions were only differentiated by a metadata field in the report. Consequently, an adversary could downgrade the firmware, obtain code execution on the ASP to extract the CEK, and forge reports for any version\cite{buhren2019insecure}. To address this, AMD introduced the \ac{vcek}\cite{VCEK-spec,amd-aspfw} alongside the SEV-SNP architecture. Instead of exposing the root seed, the VCEK mechanism is supposed to provide the ASP firmware with a derived key. This key is the product of sequentially hashing the root seed via a one-way "hashstick," where the number of hash iterations is inversely correlated with the firmware's Security Version Number (SVN). Consequently, the one-way property of the hash function ensures that older firmware cannot derive the cryptographic keys of newer versions.

Under this architecture, safeguarding the root VCEK seed is of paramount importance. If an adversary successfully extract this root seed, the entire anti-rollback mechanism would be rendered useless, as they could trivially forge attestation reports for any updated firmware version. However, AMD provides no public documentation regarding the mechanisms protecting this seed; both the ASP and the underlying processor hardware operate as completely proprietary black boxes. This raises a critical question:

\begin{quote}
\textit{Is the VCEK root seed protection securely implemented?}
\end{quote}

In this paper, we answer this question in the negative at least for the first-generation SEV-SNP processors (EPYC Milan) by demonstrating a software-only root VCEK extraction attack. 

Specifically, the foundational step of our attack requires achieving software-only arbitrary code execution within the Milan ASP firmware stack. As no prior work has demonstrated such a capability, we introduce a novel exploit methodology termed the MilanLaunchy attack. This technique targets a firmware decryption flaw in the Milan ASP BootROM. While this flaw was briefly acknowledged in an official security 
bulletin in 2021, it completely lacked technical details and its practical exploitability remained unknown. We achieved the first successful exploitation of this vulnerability, enabling us to load and execute arbitrary custom firmware.

Second, equipped with code execution on an older firmware version, we deeply reverse-engineered the protection mechanisms securing the VCEK root seed. We discovered that although AMD applies read protections to security-critical fuses, they failed to correctly implement write protections. Based on this architectural oversight, we propose the BadFuse attack, demonstrating that an adversary with code execution on an outdated firmware can successfully extract the root VCEK seed. We detail two distinct extraction methods. The first technique enables us to burn a custom public key into the silicon, allowing the platform to load and execute a malicious firmware image configured with a higher SVN. The second technique constructs a side-channel oracle to recover the root seed bit by bit. Executing either of these independent methods successfully yields the VCEK root seed.

By chaining the MilanLaunchy and BadFuse attacks, we accomplish the complete, software-only extraction of the root VCEK seed. We have responsibly disclosed all our findings and vulnerabilities to AMD.

\vspace{1em}

\textbf{Contributions.} Our key contributions are as follows:
\begin{itemize}
    \item We introduce the MilanLaunchy attack, which is the first software-based exploit achieving code execution on an legacy ASP firmware.

    \item We introduce the BadFuse attack, demonstrating how code execution within the ASP enables the extraction of the hardware root seed and subsequently forge attestation reports.

    \item We responsibly disclosed the attack chain to AMD and discussed potential mitigation strategies.
    \end{itemize}

\textbf{Responsible Disclosure.} We responsibly disclosed the MilanLaunchy attack to the AMD on January 6, 2026. AMD acknowledged the vulnerability, requested an embargo until May 12, 2026, and issued a corresponding security brief\cite{amd-3045}. 
We responsibly disclosed the BadFuse attack to the AMD on April 8, 2026. The official AMD response concerning this specific vulnerability is provided in Appendix \ref{response}. 
A detailed source code for a complete end-to-end exploit is publicly available at \url{https://github.com/muyan29}.

\section{Background\label{sec:background}}
\subsection{AMD SEV-SNP}
To address the growing security demands of cloud computing, AMD introduced its \ac{sev}\cite{AMDSevAPI} architecture across successive EPYC processor generations. The foundational Naples and Rome series introduced initial SEV support and its enhanced version SEV-ES. Specifically, the original SEV ensures memory confidentiality by transparently encrypting VM memory using ephemeral keys, while SEV-ES extends this protection by encrypting CPU register states during VM Exits, preventing a compromised hypervisor from reading or altering the VM's execution state. While SEV and SEV-ES provide strong confidentiality, they did not fully guarantee memory integrity. A malicious hypervisor could still launch advanced attacks by manipulating the nested page tables (e.g., remapping memory, replaying old data, or creating memory aliases). 

Starting with the third-generation EPYC architecture, Milan, AMD processors begin supporting the SEV-SNP feature\cite{sev2020strengthening}. SEV-SNP introduces the Reverse Map Table (RMP), a hardware-enforced data structure that strictly tracks and verifies the ownership of every physical memory page, providing strict hypervisor isolation to guarantee memory integrity. Subsequent architectures have maintained the foundational SEV-SNP framework while integrating additional security features like Trusted I/O and advanced hardware capabilities like segmented RMP, and ciphertext hiding\cite{SEV_HOME}.

\noindent \textbf{SEV-SNP Remote Attestation.} 
To verify that the VM is genuinely secure, SEV-SNP utilizes a hardware-rooted remote attestation mechanism. During the boot process, the guest VM initiates a request to the ASP via the \texttt{SNP\_GUEST\_REQUEST} Application Binary Interface (ABI)\cite{AMDSNPAPI158} command. In response, the ASP generates a cryptographically signed attestation report that encapsulates the VM's precise configuration, initial memory measurement, and the system's TCB version. This report is authenticated using either the \ac{vcek}\cite{VCEK-spec} or the Versioned Loaded Endorsement Key (VLEK)\cite{AMDSNPAPI158}. The VCEK is a machine-unique ECDSA key pair derived from the hardware's unique secret and its TCB version; we provide a comprehensive analysis of its derivation in Section \ref{sec:vcekintro}. The VLEK utilizes the VCEK as a secure transport channel to enable \acp{csp} to deploy customized keys, thereby inheriting the same foundational security guarantees as the VCEK.

To trust this report, the guest owner verifies the hardware signature against a certificate chain that is anchored to the AMD Root Key (ARK). These certificates are securely fetched and distributed via the AMD Key Distribution Service (KDS). By validating the attestation report using the KDS-provided certificate chain, the guest owner can cryptographically prove the VM is running in a pristine state on genuine AMD hardware before provisioning it with sensitive decryption keys or data.

\noindent \textbf{SEV-SNP Firmware Updates.} Across successive EPYC processor generations, AMD proactively addresses security vulnerabilities and delivers feature enhancements through continuous firmware updates that encompass both the ASP firmware stack and x86 CPU microcode. These updates can be deployed via two distinct mechanisms: cold updates and hot (live) updates. Cold updates involve updating the BIOS firmware image stored in the motherboard's SPI-Flash memory, which automatically loads the new ASP firmware and microcode during the subsequent system boot sequence. Alternatively, hot updates leverage the dedicated \texttt{SNP\_DOWNLOAD\_FIRMWARE(\_EX)} ABI to dynamically inject new SEV-SNP firmware\cite{using-snp-guide}, or concurrently utilize specialized microcode update instructions to patch the CPU microcode after the system has fully booted.

\subsection{AMD Secure Processor}
The ASP, formerly known as the Platform Security Processor (PSP), serves as the hardware-based root of trust for AMD EPYC platforms. It is an integrated ARM-based microcontroller that operates independently from the main x86 cores. In the context of the SEV-SNP architecture, the ASP is the central orchestrator responsible for managing the lifecycle of Confidential Virtual Machines (CVMs). Its primary duties include the implementation of the ABI, which encompasses critical security operations such as generating hardware-backed remote attestation reports, managing memory encryption keys, and enforcing guest-host isolation.

To manage various on-chip hardware, the ASP employs the System Management Network (SMN) as its primary communication fabric, allowing it to interface with the external SPI-Flash, control memory controllers, and access internal fuses.
The ASP also includes a Cryptographic Co-processor (CCP) to ensure high-performance security operations, executing cryptographic tasks such as AES encryption, SHA hashing, and RSA signatures\cite{specter-2024-part2}. Additionally, a Local Storage Block (LSB) is built into the CCP, providing a secure, internal memory region used to store sensitive data such as active encryption keys.

\noindent \textbf{Boot Process.}
The security of the ASP is rooted in a multi-stage boot sequence that establishes a verifiable chain of trust. The process originates from the on-chip BootROM, which serves as the immutable root of trust. Operating at the EL1 privilege level, the BootROM’s initial task is to locate, authenticate, and load the off-chip bootloader from the external SPI-Flash. Furthermore, the BootROM is responsible for facilitating other security-critical operations associated with the on-chip fuses.

Once the off-chip bootloader is verified and executed, it proceeds to load and authenticate various AMD Generic Encapsulated Software Architecture (AGESA) modules. These modules are responsible for the fundamental initialization of the system hardware, including memory training. Following the completion of these tasks, the memory space previously occupied by these modules is overwritten by the final SEV-SNP firmware.
In its final operational state, the SEV-SNP firmware runs at the unprivileged EL0 level, communicating with the resident privileged bootloader via Supervisor Calls (SVCs) to execute these sensitive operations.
\begin{figure}[tbp]
  \includegraphics[width=1\columnwidth]{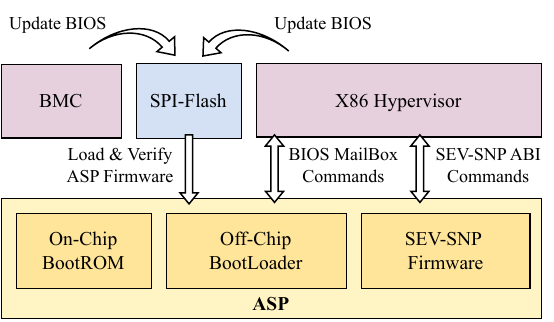}

  \caption{The software-based attack surface of the ASP.}
  \label{fig:asp-attack-surface}

\end{figure}

\noindent \textbf{Attack Surface.}
The ASP presents a complex attack surface that can be categorized into software-based and physical attack vectors.
As illustrated in Figure \ref{fig:asp-attack-surface}, the software-based threats primarily revolve around two major interfaces\cite{cohen2022amd}:

The first is the security of the firmware loading process. The ASP retrieves its firmware and microcode from the BIOS image residing in the SPI-Flash, which creates a potential attack vector for adversaries who gain write access to the flash memory. For instance, an attacker could attempt to flash a malicious BIOS image via the x86 UEFI Shell or the \ac{bmc} while the system is powered down. Prior research has demonstrated that insecure firmware loading sequences on Zen 1 processors could lead to the ASP executing unauthorized firmware code that lacks a valid AMD signature\cite{CTSLABS2018,buhren_uncover_2019}.

The second is the security of communication interfaces. Post-initialization, the ASP maintains several active interfaces with the x86 host. These include the BIOS Mailbox, used for handling platform events such as S3 sleep and wake transitions, and the SEV-SNP command interface, which processes management commands issued by the hypervisor. This area has similarly been the focus of prior security studies\cite{CTSLABS2018,tom2024blackhat}.

Existing physical attacks primarily focus on fault injection techniques, such as voltage or clock glitching. Research has demonstrated that precisely timed fault injections can be used to bypass signature verification checks during the ASP's boot process, allowing the loading of malicious firmware\cite{buhren2021one,jacob2023faultpm}. While these attacks highlight significant hardware vulnerabilities, it is important to note that physical access attacks are generally considered out-of-scope for the standard security model of confidential computing\cite{sev2020strengthening}, which primarily aims to protect against malicious software.

\subsection{VCEK Architecture in AMD SEV-SNP}
\label{sec:vcekintro}

In SEV-SNP, the endorsement key is specifically known as the \ac{vcek}. Its version designation reflects the TCB version, which primarily consists of the firmware levels running on the ASP. 
To remediate emerging threats, AMD issues periodic ASP firmware updates, which in turn increment the TCB version reflected in the attestation process.

\noindent \textbf{Minimizing Access to the Root Seed.}
Similar to the \ac{dice}\cite{tcg_dice} architecture, the VCEK mechanism in SEV-SNP is supposed to isolate the root seed to the BootROM. Notably, the lack of this isolation in earlier iterations (SEV and SEV-ES) enabled practical attacks. In those versions, the static CEK remained accessible to subsequent boot stages. Consequently, adversaries could extract the CEK or the underlying root seed from a compromised platform. Because the CEK was statically bound and could not be rotated, firmware updates were ineffective in mitigating such breaches.

\begin{figure}[tbp]
  \includegraphics[width=1\columnwidth]{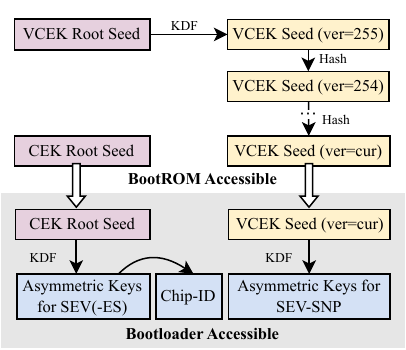}

  \caption{A comparison of key derivation flows for CEK and VCEK (simplified).}
  \label{fig:vcek-cmp}

\end{figure}

As illustrated in Figure \ref{fig:vcek-cmp}, the VCEK mechanism exposes only a derived value to the off-chip bootloader. This is achieved by applying a specific number of hash iterations to the root seed. This derivation process cryptographically guarantees that obtaining an older version of the VCEK seed poses no security threat to newer VCEK seed versions.
It is worth noting that SEV-SNP-capable EPYC processors retain the CEK to maintain compatibility with legacy SEV/SEV-ES virtual machines. Furthermore, the public key derived from the static CEK acts as the Chip-ID. This identifier allows verifiers to fetch the appropriate VCEK-signed certificate from the KDS based on the platform's TCB version.

\noindent \textbf{VCEK Layering Architecture.} 
In the AMD EPYC Milan SEV-SNP implementation, the operational TCB hierarchy is structured into four distinct layers:
\begin{itemize}
\item \textbf{Layer 0:} On-chip BootROM (ASP, EL1).\vspace{-0.5em}
\item \textbf{Layer 1:} Off-chip Bootloader (ASP, EL1).\vspace{-0.5em}
\item \textbf{Layer 2:} SEV Firmware (ASP, EL0).\vspace{-0.5em}
\item \textbf{Layer 3:} CPU Microcode (x86).
\end{itemize}

While the Layer 0 BootROM remains immutable post fabrication, Layers 1 through 3 are each characterized by an 8 bit Security Version Number (SVN) ranging from 0 to 255. These individual SVNs are aggregated to define the platform TCB version, which is subsequently used to request the corresponding VCEK certificate.

Algorithm \ref{alg:vcek_algx} details the per layer seed derivation process within the firmware architecture.
This hash chain architecture cryptographically ensures that possessing a seed from an older version cannot lead to the derivation of a newer version's seed. As an inherent byproduct of this design, a newer seed can deterministically compute the seeds of older versions. This specific property is intentionally leveraged to facilitate TCB Rollback attestation support\cite{AMDSNPAPI158}. During the cross layer transfer of the hash chain, a constant zero padding is prepended to the input. This structure ensures that the final output seed is distinct from the versioning chain, preventing cross layer interference.

\label{vcek-der}

\newcommand{\pk}{\mathit{pk}}
\newcommand{\sk}{\mathit{sk}}
\newcommand{\KDSSeed}[1]{\mathit{KDSSeed}_{#1}}
\newcommand{\MasterSeed}{\mathit{MasterSeed}}
\newcommand{\FW}{\mathit{FW}}
\newcommand{\FWBody}{\mathit{FWBody}}
\newcommand{\Ver}{\mathit{Ver}}
\newcommand{\SeedVer}[1]{\mathit{Seed\_Ver}_{#1}}
\newcommand{\VCEK}{\mathit{VCEK}}
\newcommand{\maxsvn}{\mathit{MAX\_SVN}}
\newcommand{\Sign}{\text{Sign}}

\begin{algorithm}[tbp]
\caption{VCEK-Style Per-Layer Seed Derivation}
\label{alg:vcek_algx}
\begin{algorithmic}[1] %
\Require Seed for Layer $l$ $SeedLayer_l$ or VCEK Root Seed $RootSeed$, Layer $l+1$ Firmware SVN $cur$;
\Ensure Seed for Layer $l+1$ $SeedLayer_{l+1}$;
\If{$l=0$}
\State  $TmpSeed_{255} \gets RootSeed$
\Else
\State $TmpSeed_{255} \gets SeedLayer_l$
\EndIf
    \For{$i \gets 255$ \textbf{down to} $0$}
    \State $TmpSeed_{i-1} \gets \text{Hash}(TmpSeed_{i})$
\EndFor
\If{$l=0$}
\State  Lock $RootSeed$, Erase $ \{TmpSeed_{i} \mid cur < i \le  255\}$
\EndIf
\State \text{Transfer} $TmpSeed_{cur-1}$ to Layer $l+1$ \text{for TCB Rollback}

\State \Return \hspace{-0.3em} $SeedLayer_{l+1} \hspace{-0.2em} := \hspace{-0.1em} \text{Hash}('00000000'||TmpSeed_{cur})$

\end{algorithmic}
\end{algorithm}

\section{Threat Model}
Confidential computing aims to completely decouple the host's privileged software stack (e.g., hypervisor) from the guest's TCB via hardware-backed isolation mechanisms.
Our adversary model strictly aligns with the standard confidential computing threat model\cite{sev2020strengthening}.  
Specifically, we assume an adversary with the following capabilities:
\begin{enumerate}
    \item The adversary possesses complete root-level control over the host environment. This includes the assumption that the hypervisor and host BIOS are fully untrusted and potentially malicious.
    \item The adversary has the capability to arbitrarily update, downgrade, or modify the firmware contents stored in the SPI-Flash. In practical server environments, this can be achieved via software-accessible interfaces such as the BMC or the UEFI Shell. 
\end{enumerate}

\section{Experimental Setup}
We conducted our experiments (including the MilanLaunchy and the BadFuse attack) on AMD EPYC 7413 (Milan) processors paired with a TYAN S8036GM2NE motherboard. To flash the SPI-Flash firmware, we utilized the onboard BMC. Our software environment consisted of Ubuntu 22.04 running Linux kernel version 6.12.

\section{MilanLaunchy Attack}
\label{sec:milanlaunchy}
The first step of our attack methodology requires arbitrary code execution on any accessible firmware version, including outdated legacy releases. Given the absence of known, purely software based code execution vulnerabilities on EPYC Milan, our primary strategy is to mine historical AMD security bulletins for potential ASP vulnerabilities\cite{amd-1021,amd-3002,amd-3014}. 
It is worth noting that unlike flaws disclosed at academic venues, most vulnerabilities listed in these security bulletins
lack actionable technical details. Therefore, identifying a viable attack vector demands extensive reverse engineering of the target firmware to extract the exact root cause and determine if the underlying flaw can be leveraged into a stable, real world exploit.

In this section, we first present our comprehensive analysis including reverse engineering of a firmware decryption vulnerability (CVE-2021-26315\cite{amd-1021}) disclosed for EPYC Milan in 2021 (Section \ref{sec:milanlaunchy-p1}). Subsequently, we introduce a novel exploit that achieves stable arbitrary code execution within the ASP without disrupting the normal boot sequence of the x86 operating system (Section \ref{sec:milanlaunchy-p2}). Finally, we evaluate the security mitigations deployed by AMD in 2021 to address this flaw (Section \ref{sec:milanlaunchy-p3}).

\subsection{Flaw Investigation}
\label{sec:milanlaunchy-p1}

\begin{table*}[t]
\centering
\caption{Comparison of ASP Encrypted Firmware Authentication Schemes across AMD Architectures.}
\label{tab:asp_firmware_auth}
\begin{tabular}{@{}l l l c@{}}
\toprule
\textbf{Architecture} & \textbf{WP\_IKEK Auth} & \textbf{Firmware Auth}  & \textbf{Vulnerable} \\
\midrule
Zen 1-style  & None & $\{\mathit{Header} \parallel \text{Decrypt}(\mathit{Body}) \}\text{ with RSA-PSS}$  & No \\
\addlinespace
Zen 3-style (Milan)  & $\text{with HMAC}$ & $\{\mathit{Header} \parallel \mathit{Body} \}\text{ with RSA-PSS}$ & Yes \\
\addlinespace
Zen 4/5-style  & $\text{with HMAC}$ & $\{\mathit{Header} \parallel \mathit{Body} \}\text{ with RSA-PSS} \text{ + } \{\text{Decrypt}(\mathit{Body}) \}\text{ with SHA256} $ & No \\
\bottomrule
\end{tabular}
\end{table*}

\begin{figure}[tbp]
\centering
  \includegraphics[width=1\columnwidth]{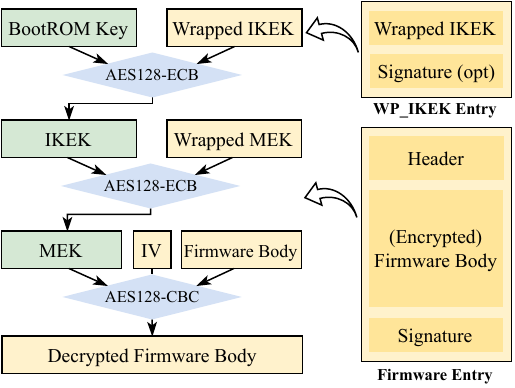}

  \caption{The specific cryptographic loading and decryption workflow for encrypted ASP firmware residing in the SPI-Flash.}
  \label{fig:encfw}

\end{figure}

\noindent \textbf{ASP Firmware Decryption Architecture.}
Figure \ref{fig:encfw} illustrates the hierarchical decryption workflow for ASP firmware, detailing the derivation and utilization of the Intermediate Key Encryption Key (IKEK) and the Module Encryption Key (MEK)\cite{psptool-github,specter-2024-part2}.

The root of this decryption chain relies on the BootROM Key, which is stored within the BootROM itself. During early system initialization, this key is loaded into the Local Storage Block (LSB) of the CCP to facilitate subsequent decryption operations. The encrypted counterpart to the first-stage key, known as the Wrapped IKEK, resides in the off-chip SPI-Flash and is retrieved from the ASP firmware directory via entry type 0x21. During the BootROM execution phase, the ASP utilizes the BootROM Key to decrypt the Wrapped IKEK via AES-128-ECB, yielding the plaintext IKEK. This intermediate key is then temporarily cached in the secure ASP SRAM.

The second stage of the pipeline occurs whenever the ASP needs to load a specific encrypted firmware module. The boot sequence first parses the firmware's header to extract the Wrapped MEK alongside its corresponding Initialization Vector (IV). The ASP then uses the previously cached IKEK to decrypt the Wrapped MEK via AES-128-ECB, successfully recovering the plaintext MEK. Finally, the derived MEK and the IV are used in an AES-128-CBC operation to decrypt the encrypted firmware body, producing the fully decrypted firmware ready for execution.

\noindent \textbf{ASP Firmware Authentication Mechanisms.}
Although AMD employs an identical decryption methodology for encrypted firmware across various processor architectures including both the Ryzen and EPYC series, our reverse engineering reveals that the authentication mechanisms applied to the decrypted payload vary significantly across generations. Crucially, the approach implemented on the Milan architecture introduces a severe vulnerability. Table \ref{tab:asp_firmware_auth} illustrates a comparative analysis of encrypted firmware samples across these architectures.
In the Zen 1 ASP architecture, the cryptographic signature is computed over the combination of the header and the decrypted firmware body. Although there is no explicit authentication for the Wrapped IKEK entry, it remains structurally secure because any tampering with the Wrapped IKEK inherently alters the resulting decrypted firmware. Since the RSA signature strictly validates the combination of the header and the decrypted payload, such modifications inevitably cause the verification process to fail.

Conversely, on the Milan architecture, the target of the RSA-PSS signature shifted to the combination of the header and the original encrypted firmware body. This architectural decision is fundamentally flawed. If adversaries can manipulate the decryption key, they can arbitrarily alter the final decryption output, coercing the ASP into executing unintended code. It is worth noting that Milan introduces an HMAC validation for the Wrapped IKEK. However, this addition fails to mitigate the underlying vulnerability. Because HMAC relies on a symmetric algorithm, the identical key is utilized for both tag generation and verification. Furthermore, since the firmware is uniformly distributed, this BootROM Key remains identical across all Milan machines. Consequently, if we manage to extract this BootROM Key from any Milan processor by any means, we can compromise the decryption key and ultimately control the decrypted payload.

In the more recent Zen 4 and Zen 5 architectures, while the signature still covers the header and the encrypted body, AMD introduced an additional integrity validation step. Specifically, the processor computes the hash SHA256 or SHA384 of the decrypted firmware and compares it against a reference hash stored directly within the header. Because the integrity of the header itself is strictly guaranteed by the RSA PSS signature, this comprehensive scheme effectively secures the decrypted payload.

\subsection{Exploit Development}
\label{sec:milanlaunchy-p2}
\noindent \textbf{Analyzing the Exploitable Execution Space.}
Decrypting valid firmware using an arbitrary incorrect key inherently yields non executable garbage data. However, by meticulously searching the decryption key space, we can deterministically control a specific number of bits within the decrypted output. Specifically, if we treat the multiple AES rounds entirely as a random permutation, controlling the initial $n$ bits of the decrypted firmware requires a brute force complexity of approximately $2^n$ operations.

In the ARM execution state, a standard instruction occupies exactly 4 bytes. Consequently, dictating the very first executed instruction necessitates a brute force effort on the order of $2^{32}$ operations. This represents a highly practical and entirely acceptable computational threshold. In stark contrast, attempting to control a second consecutive instruction escalates the complexity to an infeasible $2^{64}$ operations. In practice, this search process averages only a few minutes to complete.

Fortunately, a 4 byte execution window is fully sufficient to harbor a single branch instruction. This provides the necessary initial foothold to redirect the execution flow to the actual payload sequence, thereby achieving full arbitrary code execution. Within the ARM architecture, candidate instructions for this pivotal jump include b, bl, and bx.

\begin{figure}[tbp]
  \includegraphics[width=1\columnwidth]{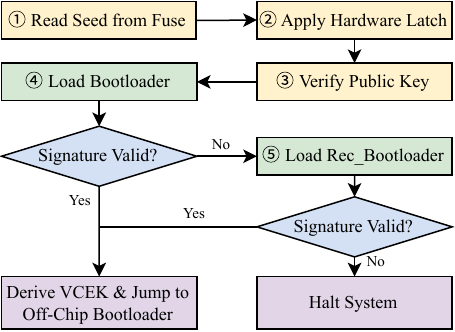}

  \caption{The BootROM execution sequence illustrating the off-chip bootloader verification and VCEK derivation.}
  \label{fig:load-off-chip}

\end{figure}

\noindent \textbf{Payload Injection.}
To introduce a sufficiently large and controllable payload, we exploit the recovery mechanism utilized by the BootROM when loading the off-chip bootloader. As illustrated in Figure \ref{fig:load-off-chip}, our reverse engineering reveals that after completing the fuse preprocessing, the BootROM initially attempts to load the primary bootloader identified by entry type 0x1. It copies the firmware content into the internal SRAM based on the load address and firmware size specified in the header. Subsequently, it performs signature verification and on demand decryption. If any of these verification steps fail, the BootROM automatically attempts to load the fallback recovery bootloader corresponding to entry type 0x3 using an identical procedure.

A critical observation is that even if the signature verification for the primary bootloader fails, the content previously copied to the SRAM is never cleared. This architectural oversight provides the exact primitive required to inject arbitrary code into the SRAM.

Finally, we construct a malicious firmware image by carefully combining an invalid primary bootloader with a legitimate recovery bootloader. We begin by replacing the standard bootloader with a specially crafted image and deliberately modify its target load address to 0x20000. This specific offset ensures that our injected data remains stale and persists in the SRAM without being overwritten by the subsequent loading of the recovery bootloader. At the very beginning of this crafted image, we embed an extra loader designed to relocate the full custom bootloader to its proper execution address. Correspondingly, the metadata of signed firmware size and image size fields are adjusted to accommodate this added payload. Next, we replace the recovery bootloader with an extracted version from Milan-PI-1.0.0.1. Crucially, this legacy version retains a valid AMD signature, allowing it to successfully pass the initial BootROM verification. Finally, the Wrapped IKEK entry is overwritten with the precise collision value calculated during our earlier brute force key search phase.

\begin{figure}[tbp]
  \includegraphics[width=1\columnwidth]{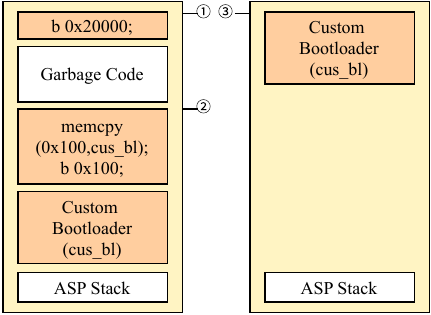}

  \caption{The memory layout and PC trajectory within the ASP SRAM during the vulnerability exploitation.}
  \label{fig:sram}

\end{figure}

\noindent \textbf{Custom Bootloader Design.}
Our exploit is strategically designed to execute a complete custom bootloader capable of fully booting the underlying x86 operating system. As shown in Figure \ref{fig:sram}, our methodology stably loads this custom bootloader as if it were a legitimate cryptographic image, it preserves the integrity of the ASP stack avoiding the corruption issues prevalent in prior Zen 1 stack overflow exploits\cite{buhren_uncover_2019}. Furthermore, it ensures all critical initialization sequences are executed compared with the incomplete state limitations typical of fault injection attacks\cite{buhren2021one}. This unprecedented stability guarantees a smooth x86 system boot. Consequently, we can execute the actual payload on the ASP for instance, extracting internal hardware secrets and seamlessly write the resulting output directly into the x86 DRAM. The operating system can then read these compromised values natively from memory, entirely eliminating the reliance on physical hardware monitors like SPI-Flash bus analyzer. This capability solidifies our methodology as a purely software based end to end attack.

In our specific instantiation, the custom bootloader is engineered to mirror the exact functionality of the standard bootloader, with one vital modification: it hooks the SEV-SNP firmware loading sequence to authorize the execution of arbitrary SEV-SNP firmware. This architectural choice grants us the tactical flexibility to embed our ultimate payloads directly within the SEV-SNP firmware image. We can then leverage the standard Linux SEV-SNP firmware management interface to dynamically load and execute the targeted payloads on demand.

\subsection{Mitigation Evaluation}
\label{sec:milanlaunchy-p3}
Because the ASP BootROM is physically immutable and the Milan architecture lacks an anti-rollback mechanism for the off-chip bootloader, the foundational vulnerability remains permanently exploitable. An adversary can reliably downgrade to the legacy firmware and trigger the exploit at will. To neutralize this threat within the SEV-SNP trust model, AMD deployed a critical update beginning with Milan-PI-1.0.0.3. Specifically, our reverse engineering found that they ceased the distribution of encrypted firmware payloads and concurrently incremented the SVN of the off-chip bootloader, aiming to cryptographically isolate the vulnerable execution state.

From a cryptographic perspective, this mitigation strategy is effective. While our method still successfully grants arbitrary code execution on the downgraded firmware, the attack is intrinsically bound to the security state of that specific legacy version. The exploit operates by hijacking the initial instruction execution such as an unconditional branch to 0x20000 immediately after the deliberate misdecryption of the lowest SVN firmware. Consequently, as shown in Figure \ref{fig:load-off-chip}, the hardware derives a VCEK seed that is tied to this minimum SVN. Due to the irreversible nature of the hardware derivation hash functions, attackers are cryptographically trapped and cannot compute the VCEK seeds corresponding to any updated firmware versions, as detailed in Figure \ref{fig:vcek-cmp}.

\section{BadFuse Attack}
This section details the BadFuse attack, a software based attack that exploits the ASP fuse controller by abusing pre-existing arbitrary code execution. If the initial code execution on the ASP is achieved via a pure software exploit (e.g. the MilanLaunchy attack in Section \ref{sec:milanlaunchy}), the complete attack chain remains entirely software only.

To illustrate this, we first present our reverse engineering of the ASP fuse space, mapping the VCEK seed storage, key security fuses, and the firmware level fuse programming interfaces (Section \ref{sec:badfuse-p1}). We then demonstrate how a lack of access control during fuse programming allows attackers to extract the VCEK seed (Section \ref{sec:badfuse-p2}). We present two independent attack methods. Either approach successfully recovers the VCEK seed, allowing us to forge arbitrary SEV-SNP attestation reports, including those representing unreleased versions.
\subsection{Reverse Engineering the Fuse Controller}
\label{sec:badfuse-p1}

\noindent \textbf{Hardware Fuse Access and Layout.}
Within the ASP architecture, the security fuses are mapped into the SMN address space from 0x5D000 to 0x5E000 for read access. Table \ref{tab:efuse-map-en} details the security critical fuse layout uncovered during our reverse engineering phase. As indicated, both the CEK and VCEK root seeds occupy exactly 32 bytes. Notably, AMD implemented an optional redundancy validation mechanism specifically for the VCEK seed. This mechanism leverages ECC bits to correct bit flips in the raw seed and employs a Fletcher-32 checksum to guarantee absolute integrity. If the hardware detects uncorrectable errors or if the Fletcher-32 validation fails, the execution aborts immediately during the BootROM phase. The activation of this redundancy feature is entirely optional and is dictated by a specific bit within the metadata security control bitmap (Table \ref{tab:efuse-bitmap}).

A widely recognized function associated with the hardware fuses is the AMD Platform Secure Boot (PSB)\cite{IOACTIVE_PSB,PSB_STATUS}. This feature allows Original Equipment Manufacturers (OEMs) to irreversibly restrict the processor to execute only BIOS or UEFI firmware cryptographically signed by that specific manufacturer. As the enforcement of PSB is primarily implemented within the off-chip bootloader and is orthogonal to our SEV-SNP attack vectors, we leave additional details regarding the PSB mechanism to Appendix \ref{psb}.

Beyond these configurations, we discovered a previously undisclosed capability which we denote as the Custom\_PK feature. During a standard secure boot sequence corresponding to Step 3 in Figure \ref{fig:load-off-chip}, the BootROM authenticates the AMD public key residing in the SPI-Flash by comparing its SHA384 digest against a hardcoded value embedded within the BootROM itself. This comparison establishes the foundational root of trust for the entire platform. However, if the Custom\_PK feature is activated meaning the fuse region designated for its SHA384 digest contains non zero data the system preemptively prioritizes this custom digest. From an adversarial standpoint, exploiting this feature enables a complete displacement of the hardware root of trust, effectively transferring it from the legitimate AMD public key to an arbitrary adversary controlled key.

\begin{table}[tbp]
\footnotesize
\centering
\caption{Fuse Memory Map on EPYC Milan}
\label{tab:efuse-map-en}
\renewcommand{\arraystretch}{1.2} %
\begin{tabularx}{\columnwidth}{@{}llX@{}}
\toprule
\textbf{Address} & \textbf{Data Size} & \textbf{Core Function \& Security Description} \\ \midrule
\texttt{0x5D044} & 32 Bytes & CEK Hardware Root Seed \\
\texttt{0x5D06C} & 3 Bits   & PSB Commit flags \\
\texttt{0x5D074} & 48 Bytes & SHA-384 Hash of Custom\_PK \\
\texttt{0x5D0A4} & 30 Bytes & Dedicated ECC for Custom\_PK \\
\texttt{0x5D0C4} & 32 Bytes & VCEK Hardware Root Seed \\
\texttt{0x5D0E4} & 20 Bytes & Dedicated ECC for VCEK Hardware Root Seed \\
\texttt{0x5D0F8} & 3 Bytes  & Metadata Security Control Bitmap\\
\texttt{0x5D0FC} & 4 Bytes  & Fletcher-32 Checksum for Custom\_PK \\
\texttt{0x5D100} & 4 Bytes  & Fletcher-32 Checksum for VCEK Seed \\
\texttt{0x5D108} & 4 Bits   & Platform Model ID for PSB \\
\texttt{0x5D109} & 8 Bits   & Platform Vendor ID for PSB \\
\texttt{0x5D10C} & 1 Bit    & PSB Enable Flag \\ \bottomrule
\end{tabularx}
\end{table}

\noindent \textbf{Hardware Fuse-Blowing Process.} 
We discovered a dedicated function within the ASP firmware responsible for fuse programming, which executes physical fuse-blowing through strict Memory-Mapped I/O (MMIO) interactions. Upon execution, it initially validates the target index and reads the shadow register at SMN address 0x5d000 to prevent redundant burning. It then polls the busy bit of the control status register at 0x5e000 to ensure the hardware is in an absolute idle state. Subsequently, it writes specific control bits to 0x5e014 and configures the target address selector in 0x5e004. The actual physical burn is triggered by precisely toggling the control bit in 0x5e004, which instructs the hardware to release an instantaneous high-voltage pulse, permanently blowing the designated fuse. Finally, the function verifies the success of the operation by checking specific hardware status bits.

We discovered that this function encapsulated within the off-chip bootloader was originally intended to execute OEM specific PSB fuse programming operations. Crucially, however, this same interface can be exploited to program other secret fuses.

\noindent \textbf{Redundancy Mechanisms in fuse Programming.}
To ensure the structural integrity and physical reliability of critical hardware configurations against silicon degradation and potential programming failures, the ASP firmware implements optional redundancy programming mechanisms during the fuse blowing process.

For extended data payloads, the firmware guarantees data fidelity by coupling Error Correction Code (ECC) encoding with a Fletcher-32 checksum. A remarkable feature of this architecture is its extremely high ECC redundancy ratio: it generates 5 bits of ECC for every 8 bits (1 byte) of payload data. For example, when provisioning a 48-byte SHA-384 Custom\_PK hash, the system simultaneously programs a proportionally massive 30-byte ECC payload (48 bytes × 5 bits/byte) alongside a 32-bit (4-byte) Fletcher checksum. This allows the hardware to not only detect integrity violations but also seamlessly correct minor physical bit-flips on the fly.

Conversely, for critical single-bit control flags, the firmware utilizes a triple modular redundancy strategy through repeated programming. Instead of relying on a single fuse, a critical logical flag is repeatedly burned into multiple, physically dispersed fuse locations. During runtime evaluation, the firmware derives the final authoritative state by executing bitwise logic operations across these redundant bits. 

\noindent \textbf{Protection Mechanisms for the Secret Fuse.}
Through reverse engineering, we confirmed that AMD correctly implemented a hardware latch defense mechanism. Specifically, as illustrated in Figure \ref{fig:load-off-chip}, immediately after the BootROM loads the secret fuse contents into SRAM, it asserts a hardware latch to block any subsequent read access to this specific memory region. This latching operation is strictly unidirectional; once engaged, it cannot be reversed without a complete system power cycle. Our experiments verify that any attempt to directly read these protected SMN addresses simply returns null bytes. This architectural design successfully prevents the off-chip bootloader from extracting the root seeds directly.

\begin{figure}[tbp]
  \includegraphics[width=1\columnwidth]{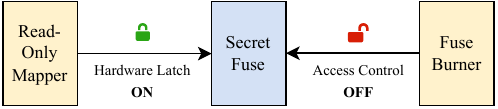}

  \caption{Secret fuse access permissions for the off-chip bootloader.}
  \label{fig:alia}
\end{figure}

However, as previously discussed, the fuse programming process utilizes discrete MMIO registers and is managed by a completely independent hardware component. Crucially, we discovered that under factory default configurations, the off-chip bootloader retains full capability to issue write operations via this dedicated fuse burner (Figure \ref{fig:alia}). This vulnerability suggests either a fundamental omission of write protection for the secret fuse region at the hardware level, or the existence of a protection mechanism that remains disabled by default. In the subsequent section, we detail how an adversary can exploit this critical absence of write protection to successfully extract the foundational VCEK root seed.

\subsection{Attack Methodology}
\label{sec:badfuse-p2}
\noindent \textbf{Attack I: Manipulating the Custom Public Key Hash.} 
This attack is based on the following observations:
\begin{enumerate}
    \item We can write the Custom\_PK fuse region, effectively transferring the foundational AMD root of trust to an adversary controlled counterpart.
    \item As depicted in Figure \ref{fig:vcek-cmp}, extracting the VCEK seed associated with an SVN of 255 is cryptographically equivalent to obtaining the absolute root VCEK seed under the existing VCEK derivation architecture.
\end{enumerate}

These two observations empower us to first overwrite the hardware public key hash with the digest of our custom public key. Since we control the corresponding private key, we can cryptographically sign a malicious firmware payload explicitly configured with an SVN of 255 and subsequently flash it onto the SPI-Flash.

Finally, this methodology allows us to extract the same seed utilizing a strictly software based attack vector, an achievement that previously required complex physical attacks as detailed in \cite{buhren2021one}. By acquiring this root seed, we can trivially derive the VCEK seed for any arbitrary firmware version by traversing the established hash chain.

\noindent \textbf{Attack II: Direct Extraction of the VCEK Seed.}
This attack is based on the following observations:
\begin{enumerate}
    \item We can do write operations to the fuse region housing the VCEK seed. Critically, we can manipulate this target fuse area with a precise 1 bit granularity.
    \item We found that attempting to re-blow a fuse bit that is already set to 1 does not brick or damage the underlying hardware.
    \item We discovered that although ECC and Fletcher-32 validation mechanisms were architecturally designed for the VCEK fuse region, they remain disabled in practice.
\end{enumerate}

These observations collectively enable us to systematically recover the entire VCEK seed bit by bit through an iterative trial writing process.

As shown in Algorithm \ref{alg:vcek_oracle}, the attack initiates by targeting a specific bit within the root VCEK fuse region. By interacting directly with the SMN controller, the attacker can blindly assert the target fuse bit to '1'.
Because this physical state alteration does not automatically update the runtime keys residing in memory, the system must undergo a power cycle. This cold reboot clears the hardware latches and forces the ASP BootROM, the first component to execute upon startup, to fetch the newly modified physical fuses and execute the complete key derivation process from scratch.

Upon successful initialization, the adversary observes the oracle's output by retrieving the newly generated VCEK. The final VCEK seed is the product of multiple cryptographic hashes originating from the root seed. If the targeted fuse bit was originally '0' and was successfully blown to '1', this single physical alteration will propagate through the derivation chain, resulting in a completely distinct final VCEK. Conversely, if the bit was already '1', the blind write operation has no physical consequence, and the newly derived VCEK will perfectly match the baseline. By evaluating the presence or absence of a change in the derived seed, the attacker can reliably deduce the physical state of the unreadable root fuses, achieving full extraction of the 256-bit secret.

\begin{algorithm}[tbp]
\caption{Bit-by-Bit VCEK Root Seed Extraction via Fuse Oracle}
\label{alg:vcek_oracle}
\begin{algorithmic}[1]
\Require $N = 256$: Length of the VCEK seed in bits
\Ensure $S$: The recovered original VCEK root seed array

\State $S \gets \text{Initialize array of size } N$
\State $V_{base} \gets \text{Fetch\_Current\_VCEK()}$ 

\For{$i \gets 0$ \textbf{to} $255$}
    \State $\text{Program\_Fuse\_Bit(i)}$ %
    \State $\text{Cold\_Reboot\_System()}$ %
    \State $V_{curr} \gets \text{Fetch\_Current\_VCEK()}$ %
    
    \If{$V_{curr} \neq V_{base}$}
        \State $S[i] \gets 0$ \Comment{Fuse was intact (0)}
        \State $V_{base} \gets V_{curr}$ %
    \Else
        \State $S[i] \gets 1$ \Comment{Fuse already blown (1)}
    \EndIf
\EndFor

\State \Return $S$
\end{algorithmic}
\end{algorithm}

\section{Mitigation}
To effectively disrupt this comprehensive attack chain, the primary defensive priority is mitigating the underlying BadFuse vulnerability.
Therefore, the mitigation strategies detailed below focus on securing the hardware fuse controller against the BadFuse attack vectors.

\noindent \textbf{Enhancing BootROM Security.}
As an immutable part post fabrication, the security of the BootROM is paramount. Under the SEV-SNP security model, particularly following the introduction of the VCEK, processors should not support custom root public key hashing functionalities. Disabling this feature effectively neutralizes BadFuse Attack I. Furthermore, the BootROM code, especially its cryptographic security modules, requires rigorous auditing to eliminate exploitable vulnerabilities similar to MilanLaunchy.

\noindent \textbf{Modifying the VCEK Derivation Mechanism.}
An alternative approach to mitigate BadFuse Attack I involves modifying how the VCEK is derived. For instance, incorporating the public key into the VCEK root secret derivation would successfully isolate different execution branches. Another potential mitigation is to adopt a DICE like derivation scheme\cite{tcg_dice}. Instead of relying solely on the SVN, this approach incorporates actual firmware content measurements into the derivation process, ensuring that any tampered or malicious firmware yields incorrect cryptographic keys. However, as discussed in \cite{buhren2021one}, this strategy introduces a structural compromise: it sacrifices the TCB rollback functionality intrinsic to SEV-SNP and prevents the deployment of multiple firmware releases under a single SVN.

\noindent \textbf{Hardware Level Defenses for Fuse Controllers.}
To address BadFuse Attack II, hardware designers must implement dedicated write protection control bits specifically for the VCEK fuse arrays to prevent any post fabrication modifications. Another viable defense mechanism is the mandatory enforcement of ECC and Fletcher-32 validation. Activating these error correction features and enforcing strict plaintext binding limits the number of bits an adversary can successfully manipulate. Introducing excessive errors during an attack would be definitively detected by the ECC and Fletcher-32 validation, thereby thwarting the complete extraction of the VCEK seed.

\section{Related Work\label{sec:related_work}}

\noindent \textbf{Software Attacks on the Hypervisor-Guest Boundary.} The interface between the untrusted hypervisor and the confidential guest has historically been a primary attack surface. Early implementations like SEV and SEV-ES suffered from incomplete state protection, allowing malicious hypervisors to manipulate execution flows or abuse nested page tables to corrupt guest data \cite{hetzelt2017security,morbitzer2021severity}. In SEV-SNP, these issues are effectively addressed by the RMP table. However, as the hypervisor retains partial control over the VM's nested page tables, adversaries can potentially single-step the execution of the VM using external APIC timers (e.g. SEV-Step \cite{wilke2023sev}). Furthermore, as the untrusted hypervisor can inject interrupts into the guest VM during runtime, an attacker can maliciously exploit the interrupt handlers within the guest kernel to compromise the integrity of the confidential VM, as seen in the Heckler \cite{schluter2024heckler} and WESEE \cite{schluter2024wesee} attacks. More recently, Takekoshi et al. \cite{takekoshi2025badaml} demonstrated that because ACPI tables are excluded from static boot attestation, a malicious host can leverage ACPI Machine Language (AML) to inject arbitrary code into the guest kernel. To mitigate these threats, a fundamental approach is to leverage Virtual Machine Privilege Levels (VMPLs) to enforce finer-grained isolation within the guest VM\cite{zhou2024verismo,SVSM}.

Another active line of research focuses on side-channel vulnerabilities. Specifically, the hypervisor can exploit the XEX encryption mode of guest memory to mount ciphertext side-channel attacks \cite{li2021cipherleaks,li2022systematic,yuan2025ciphersteal,schluter2025heracles,yan2025relocate}. Because this mode deterministically encrypts the same plaintext at the same physical address into identical ciphertext, adversaries can monitor memory write patterns to recover partial or complete confidential information. Similarly, attackers can leverage CPU performance counters and power monitoring interfaces to extract cryptographic secrets \cite{gast2025counterseveillance,wang2023pwrleak}.
To mitigate both ciphertext and performance counter side-channel attacks, AMD has introduced Ciphertext Hiding and Performance Counter (PMC) Virtualization technologies in its recent architectures\cite{SEV_HOME}.

\noindent \textbf{Software Attacks on ASP Firmware and X86 Microcode.}
Vendor-specific implementations, primarily the ASP firmware and x86 microcode on AMD processors, constitute another critical attack surface. Prior to the initial release of SEV-SNP, an early security audit of the ASP firmware conducted by AMD and Google revealed multiple vulnerabilities \cite{cohen2022amd}. Following AMD's partial open-sourcing of the ASP firmware in 2023\cite{amd-aspfw}, Dohrmann identified several implementation-level flaws, including typographical errors and insufficient input validation\cite{amd-sev-vul2024}. More recently, sophisticated vulnerabilities targeting the underlying architectural initialization have been discovered. The Rmpocalypse attack \cite{schluter2025rmpocalypse} demonstrates that an untrusted hypervisor can exploit a synchronization gap between the x86 cores and the ASP to maliciously control the initial RMP state, ultimately compromising the integrity of confidential VMs. Similarly, the FABRICKED attack \cite{schluterfabricked} manipulates AMD's Data Fabric routing during system initialization to bypass SEV-SNP hardware protections, enabling practical end-to-end attacks.

Flawed x86 microcode implementations also compromise the security of SEV-SNP. For instance, the Cachewarp attack \cite{zhang2024cachewarp} exploits faulty INVD instruction implementations, enabling attackers to roll back and subsequently corrupt the execution state of a guest VM. Furthermore, the EntrySign attack \cite{amd-microcode} directly targets insecure cryptographic implementations within the microcode update mechanism. This allows adversaries to load arbitrary microcode patches and completely subvert the microcode root of trust. More recently, the Stackwarp attack \cite{zhangstackwarp} exploited the improper synchronization of the stack engine across logical cores, leading to unauthorized control flow hijacking or data corruption within guest VMs. To mitigate these vulnerabilities, AMD has issued corresponding ASP firmware and x86 microcode updates with bumped SVNs.

\noindent \textbf{Physical Attacks on the ASP Firmware.}
Physically attacking the ASP to extract the VCEK root seed represents the most direct method to compromise SEV-SNP. Buhren et al. demonstrated the first fault injection attack against the ASP, leveraging anomalous voltage packets injected over the SVI2 bus to bypass the firmware's signature verification during the BootROM phase \cite{buhren2021one}. This manipulation allowed for the loading of a malicious ASP firmware image with the maximum SVN, consequently enabling attackers to extract the CEK or VCEK seeds across Zen 1 to Zen 3 architectures. Similar techniques have also been applied to compromise the firmware Trusted Platform Module (fTPM) on consumer-grade processors \cite{jacob2023faultpm}. Although fault injection typically requires opening the chassis and therefore falls outside the standard confidential computing threat model, it remains a highly potent physical attack vector.

\noindent \textbf{Physical Attacks on the Memory Subsystem.}
Attacks targeting the memory subsystem enable adversaries to bypass the memory controller, allowing direct physical read and write access to otherwise protected memory regions. The BadRAM attack \cite{badram} achieves this by modifying the \ac{spd} data on physical DIMMs to report double the actual memory capacity, enabling attackers to create memory aliases and access protected memory, which facilitates various subsequent exploits. Building upon this, the Battering RAM attack \cite{de2026battering} introduces dynamic runtime aliasing via a low-cost DDR4 interposer, successfully evading static boot-time alias detection. More recently, TEE.fail \cite{chuang2026tee,seto2025wiretap} utilizes physical probing of the DDR5 memory bus to monitor fluctuations in encrypted memory data. The leaked side-channel information allows attackers to extract the complete provisioning certification key on Intel TDX, and bypass ciphertext hiding mechanisms to re-enable ciphertext side-channel attacks on AMD SEV-SNP. Compared to fault injection, attacks on the memory subsystem generally require weaker adversarial capabilities, with some attack vectors even achievable via software in some cases\cite{amd-3015}. Consequently, AMD has selectively introduced mitigations to address certain aspects of these vulnerabilities.

\section{Conclusion\label{sec:conclusion}}

In this paper, we demonstrate a software-based attack chain to extract the VCEK root seed on EPYC Milan processors. We first introduce the MilanLaunchy attack, which achieves arbitrary code execution on the ASP through a novel exploitation of the firmware decryption vulnerability. Subsequently, we present the BadFuse attack, which leverages access control vulnerabilities within the fuse controller to successfully extract the VCEK seed.
Our attack highlights that protecting foundational root secrets is the absolute cornerstone of confidential computing.
Ultimately, our research demonstrates that strict cryptographic and access control models, alongside rigorous secure implementation, are critical for defending the TCB against advanced exploitation.

\bibliography{paper.bib}
\bibliographystyle{plain}

\clearpage 
\onecolumn
\appendix
\section{Appendix\label{sec:append}}

\begin{table*}[h]
\centering
\caption{Metadata Header Fuse Bit-map}
\label{tab:efuse-bitmap}
\scriptsize 
\begin{tabular}{cllllllll}
\toprule
\textbf{Address} & \multicolumn{1}{c}{\textbf{Bit 7}} & \multicolumn{1}{c}{\textbf{Bit 6}} & \multicolumn{1}{c}{\textbf{Bit 5}} & \multicolumn{1}{c}{\textbf{Bit 4}} & \multicolumn{1}{c}{\textbf{Bit 3}} & \multicolumn{1}{c}{\textbf{Bit 2}} & \multicolumn{1}{c}{\textbf{Bit 1}} & \multicolumn{1}{c}{\textbf{Bit 0}} \\ \midrule

\texttt{0x5d0F8} & 
\begin{tabular}[c]{@{}l@{}}Custom PK \\ Revocation (R2)\end{tabular} & 
\begin{tabular}[c]{@{}l@{}}PK Revocation \\ Index 3 (R2)\end{tabular} & 
\begin{tabular}[c]{@{}l@{}}PK Revocation \\ Index 2 (R2)\end{tabular} & 
\begin{tabular}[c]{@{}l@{}}PK Revocation \\ Index 1 (R2)\end{tabular} & 
\begin{tabular}[c]{@{}l@{}}Custom PK \\ Revocation (R1)\end{tabular} & 
\begin{tabular}[c]{@{}l@{}}PK Revocation \\ Index 3 (R1)\end{tabular} & 
\begin{tabular}[c]{@{}l@{}}PK Revocation \\ Index 2 (R1)\end{tabular} & 
\begin{tabular}[c]{@{}l@{}}PK Revocation \\ Index 1 (R1)\end{tabular} \\ \addlinespace

\texttt{0x5d0F9} & 
\begin{tabular}[c]{@{}l@{}}Custom PK \\ Fletcher Enable (R1)\end{tabular} & 
\begin{tabular}[c]{@{}l@{}}Custom PK \\ ECC Enable (R3)\end{tabular} & 
\begin{tabular}[c]{@{}l@{}}Custom PK \\ ECC Enable (R2)\end{tabular} & 
\begin{tabular}[c]{@{}l@{}}Custom PK \\ ECC Enable (R1)\end{tabular} & 
\begin{tabular}[c]{@{}l@{}}Custom PK \\ Revocation (R3)\end{tabular} & 
\begin{tabular}[c]{@{}l@{}}PK Revocation \\ Index 3 (R3)\end{tabular} & 
\begin{tabular}[c]{@{}l@{}}PK Revocation \\ Index 2 (R3)\end{tabular} & 
\begin{tabular}[c]{@{}l@{}}PK Revocation \\ Index 1 (R3)\end{tabular} \\ \addlinespace

\texttt{0x5d0FA} & 
\begin{tabular}[c]{@{}l@{}}VCEK Fletcher \\ Enable (R3)\end{tabular} & 
\begin{tabular}[c]{@{}l@{}}VCEK Fletcher \\ Enable (R2)\end{tabular} & 
\begin{tabular}[c]{@{}l@{}}VCEK Fletcher \\ Enable (R1)\end{tabular} & 
\begin{tabular}[c]{@{}l@{}}VCEK ECC \\ Enable (R3)\end{tabular} & 
\begin{tabular}[c]{@{}l@{}}VCEK ECC \\ Enable (R2)\end{tabular} & 
\begin{tabular}[c]{@{}l@{}}VCEK ECC \\ Enable (R1)\end{tabular} & 
\begin{tabular}[c]{@{}l@{}}Custom PK \\ Fletcher Enable (R3)\end{tabular} & 
\begin{tabular}[c]{@{}l@{}}Custom PK \\ Fletcher Enable (R2)\end{tabular} \\ \bottomrule

\multicolumn{9}{l}{* (R1), (R2), (R3) denote triple modular redundancy slots.}
\end{tabular}
\end{table*}

\twocolumn
\subsection{AMD Platform Secure Boot}
\label{psb}
\textbf{Overview and Hardware Binding.}
AMD introduced PSB\cite{IOACTIVE_PSB,PSB_STATUS} to allow OEMs to enforce a hardware-level root of trust, ensuring that a processor will only execute BIOS/UEFI firmware cryptographically signed by that specific manufacturer. Fresh out of the factory, AMD processors are shipped in an unfused state with PSB disabled. During the initial deployment, the PSB fuses can be permanently blown via specific BIOS commands. Once fused, the chip becomes cryptographically bound to the motherboard manufacturer. The blown fuses record critical data, including the OEM's ID and the hash of the public key used to verify the manufacturer's BIOS signatures. By design, this hardware lock prevents the processor from being detached and repurposed on a motherboard from a different vendor.
The core functionality of PSB is orchestrated by the off-chip bootloader executing on the ASP. During the boot sequence, the ASP inspects the x86 BIOS/UEFI Pre-EFI Initialization (PEI) firmware volumes stored in the SPI-Flash.

\noindent \textbf{Real-World Adoption and SEV-SNP Impact.}
Aside from a select few vendors selling highly customized, pre-built OEM machines, the vast majority of motherboard manufacturers leave PSB unconfigured. 
Crucially, regarding the AMD SEV-SNP threat model, whether PSB is enabled or disabled has no impact on the security guarantees of the confidential computing environment. PSB is designed to verify the integrity of the x86 BIOS/UEFI image. Because the foundational threat model of SEV-SNP explicitly assumes that the x86 BIOS, UEFI, and hypervisor are entirely untrusted and potentially malicious, the architectural security of SEV-SNP is independent of the platform's PSB configuration.

\subsection{Mitigating Potential Supply Chain Attacks from MilanLaunchy}
This section discusses the potential implications of the MilanLaunchy vulnerability on users who don't care about SEV-SNP. 

As discussed in Section \ref{sec:milanlaunchy-p3}, the specific SEV-SNP attack vector was addressed in the MilanPI-1.0.0.3 update through the VCEK mechanism, allowing SEV-SNP users to leverage remote attestation to verify their genuinely executing firmware version. However, because the BootROM is immutable and fundamentally lacks rollback protection, adversaries can still target users who operate outside the SEV-SNP threat model. By exploiting MilanLaunchy, attackers can bypass the PSB and implant arbitrary code while ensuring the underlying x86 system boots successfully. Consequently, this creates a supply chain attack surface for EPYC users who don't know SEV-SNP and Threadripper PRO 59x5WX users (the Threadripper PRO 59x5WX series shares the identical vulnerable BootROM architecture and lacks the VCEK mechanism). 

As recommended in the official AMD security brief\cite{amd-3045}, effectively mitigating this specific threat requires securing the SPI-Flash environment, which can be achieved by manually reflashing the SPI-Flash component with an official firmware image.

\subsection{Responsible Disclosure and Vendor Response}
\label{response}
\begin{itemize}
    \item January 6, 2026: We responsibly disclosed the MilanLaunchy attack to AMD via email with a PoC.
    \item January 20, 2026: AMD established the embargo date for the MilanLaunchy attack to February 24, 2026.
    \item January 30, 2026: AMD requested an embargo extension for the MilanLaunchy attack  to May 12, 2026 due to its impaction on Threadripper PRO series\cite{amd-3045}.
    \item April 8, 2026: We responsibly disclosed the BadFuse attack to AMD via email with a PoC.
    \item April 29, 2026: AMD issued their response addressing the BadFuse attack, stating the following:
    \begin{quote}
    \textit{
        The internal SME stated that he doesn’t believe any action is necessary on this ticket.}
        
\textit{To carry out the attacks, the attacker needs first exploit a prior attack vector (MilanLaunchy). This dependency was confirmed by the researcher.  MilanLaunchy itself exploits previously disclosed vulnerability.}
\textit{The overall scenario constitutes a chain attack.}
    \end{quote}
    
\end{itemize}

\end{document}